\documentclass[
prl,
twocolumn,
showpacs,
preprintnumbers,
amsmath]{revtex4}

\usepackage{graphicx}	
\usepackage{amssymb}	
\usepackage{amsmath}	
\usepackage{mathptm}	
\usepackage{color}	

\newcommand{\be}{\begin{equation}}
\newcommand{\ee}{\end{equation}}
\newcommand{\bea}{\begin{eqnarray}}
\newcommand{\eea}{\end{eqnarray}}

\begin{document}


\title{Universality of Level Spacing Distributions in Classical Chaos}

\author{J.F.~Laprise$^{a}$, J.~Kr\"{o}ger$^{b}$, 
H.~Kr\"{o}ger$^{a}$
$\footnote{Corresponding author, Email: hkroger@phy.ulaval.ca}$, 
P.Y.~St.-Louis$^{a}$, L.J.~Dub\'{e}$^{a}$, E.~Endress$^{a,c}$, A.~Burra$^{a,d}$, 
R.~Zomorrodi$^{a}$, G.~Melkonyan$^{a}$, K.J.M.~Moriarty$^{e}$}

\affiliation{
$^{a}$ {\small\sl D\'{e}partement de Physique, Universit\'{e} Laval, Qu\'{e}bec, Qu\'{e}bec G1K 7P4, Canada} \\ 
$^{b}$ {\small\sl Physics Department and Center for the Physics of Materials, McGill University, Montr\'eal, Qu\'ebec H3A 2T8, Canada} \\
$^{c}$ {\small\sl Fachbereich Physik, Universit\"at Mainz, D-55099 Mainz, Germany} \\
$^{d}$ {\small\sl Mahatma Gandhi Institute of Technology, Gandipet, Hyderabad 500075, India} \\
$^{e}$ {\small\sl Department of Mathematics, Statistics and Computing Science, Dalhousie University, Halifax, Nova Scotia B3H 3J5, Canada} 
}
\date{\today}


\begin{abstract}

We suggest that random matrix theory applied to a classical action matrix 
can be used in classical physics to distinguish chaotic from non-chaotic 
behavior. We consider the 2-D stadium billiard system as well as the 
2-D anharmonic and harmonic oscillator. By unfolding of the spectrum of 
such matrix we compute the level spacing distribution, the spectral 
auto-correlation and spectral rigidity. We observe Poissonian behavior 
in the integrable case and Wignerian behavior in the chaotic case. 
We present numerical evidence that the action matrix of the stadium 
billiard displays GOE behavior and give an explanation for it. 
The findings present evidence for universality of level 
fluctuations - known from quantum chaos - also to hold in classical physics. 

\end{abstract}


\maketitle

     

\twocolumngrid

{\it Introduction}.
Experiments and computer simulations have shown quantum systems with a fully chaotic classical counter part possess universality properties~\cite{Mehta91,Reichl92,Blumel97,Stockmann99,Haake01}. For example, the spectra originating from different heavy nuclei give the same Wignerian energy level spacing distribution~\cite{Bohigas83}. About 1950 Mehta and Wigner showed that random matrices of Gaussian orthogonal ensembles (GOE) generate a Wigner-type level spacing distribution~\cite{Mehta91}. The celebrated Bohigas-Giannoni-Schmit (BGS)
conjecture~\cite{Bohigas84} states that in time-reversal invariant quantum systems 
with fully chaotic classical counterpart, the energy level spacing 
distribution is the same as that obtained from random matrices of a certain symmetry (Gaussian orthogonal ensembles GOE), resulting in a Wignerian distribution.
This paper is about classical chaos occuring widely in nature, for example in astro physics, meteorology and dynamics of the atmosphere, fluid and ocean dynamics, climate change, chemical reactions, biology, physiology, neuroscience, or medicine. 
Traditionally, classical chaos is described by tools of nonlinear dynamics like Lyapunov exponents, Kolmogorov-Sinai entropy and phase space portraits (Poincar\'e sections).
In this work we present evidence that fully chaotic classical systems show universality behavior, visible in level spacing fluctuations obtained from a classical action matrix. 
We show that random matrix theory can be used in classical systems to distinguish chaotic from integrable systems. In the case of the 2-D billiard system, Lorentz gas and anharmonic oscillator we find GOE-type behavior of action matrix elements and Wignerian behavior in the level spacing distribution. Moreover, we provide an explanation why those action matrix elements display GOE-type behavior. This holds when the system is both, classically chaotic and diffusive (random), like the Lorentz gas and the reason is due to the Central Limit Theorem. We discuss implications on ergodicity.
\begin{figure}[ht]
\begin{center}
\includegraphics[scale=0.15,angle=0]{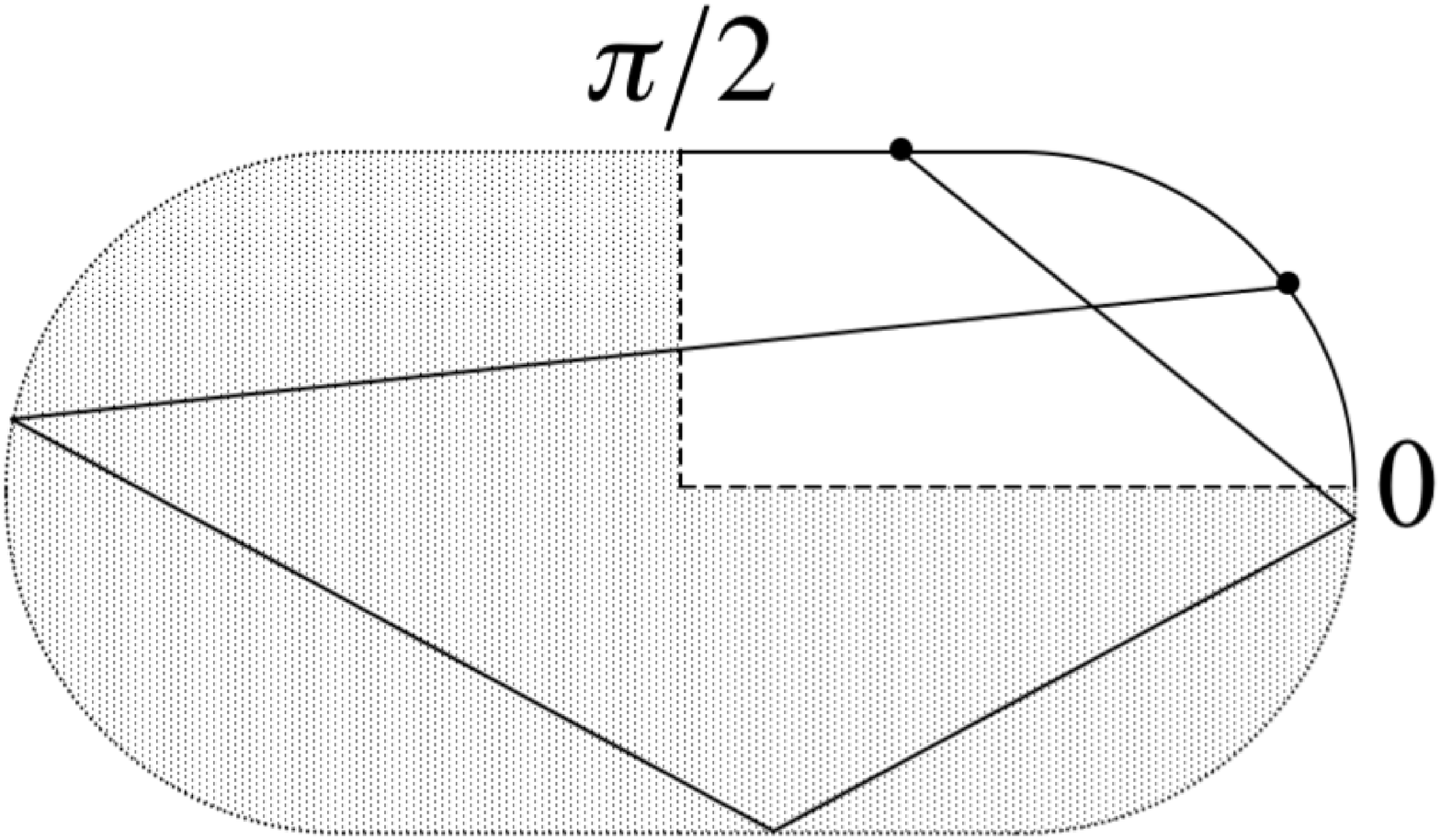}
\end{center}
\vspace{0.0cm}
\caption{
Scheme of 2-D stadium billiard system with trajectories going from $x_{i}$ to $x_{j}$ located 
on the wall of a quarter billiard. 
}
\label{fig:BilliardScheme}
\end{figure} 

{\it Classical systems: action matrix and level spacing distribution}.
\begin{figure}[ht]
\begin{center}
\includegraphics[scale=0.12,angle=0]{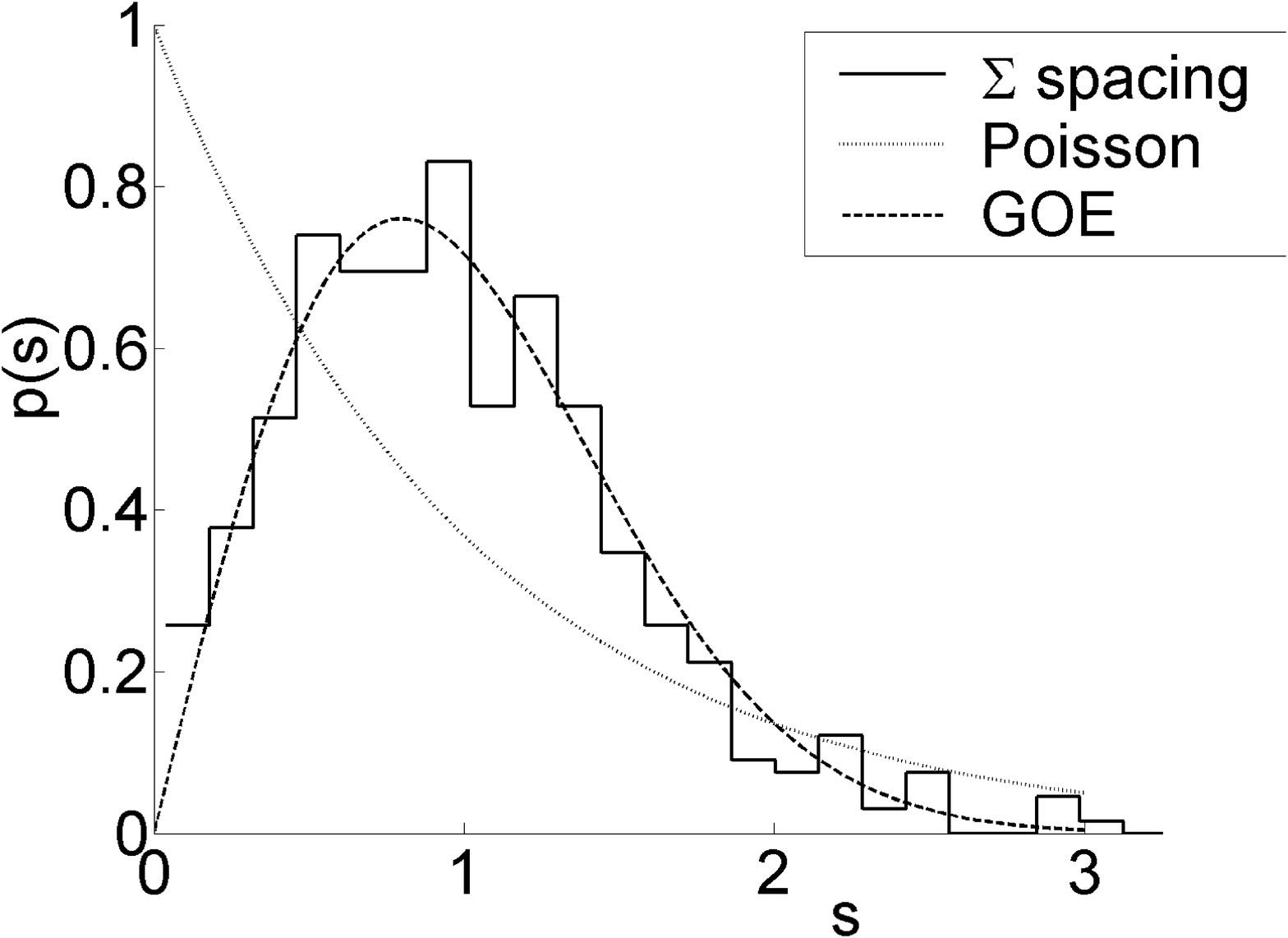}
\includegraphics[scale=0.12,angle=0]{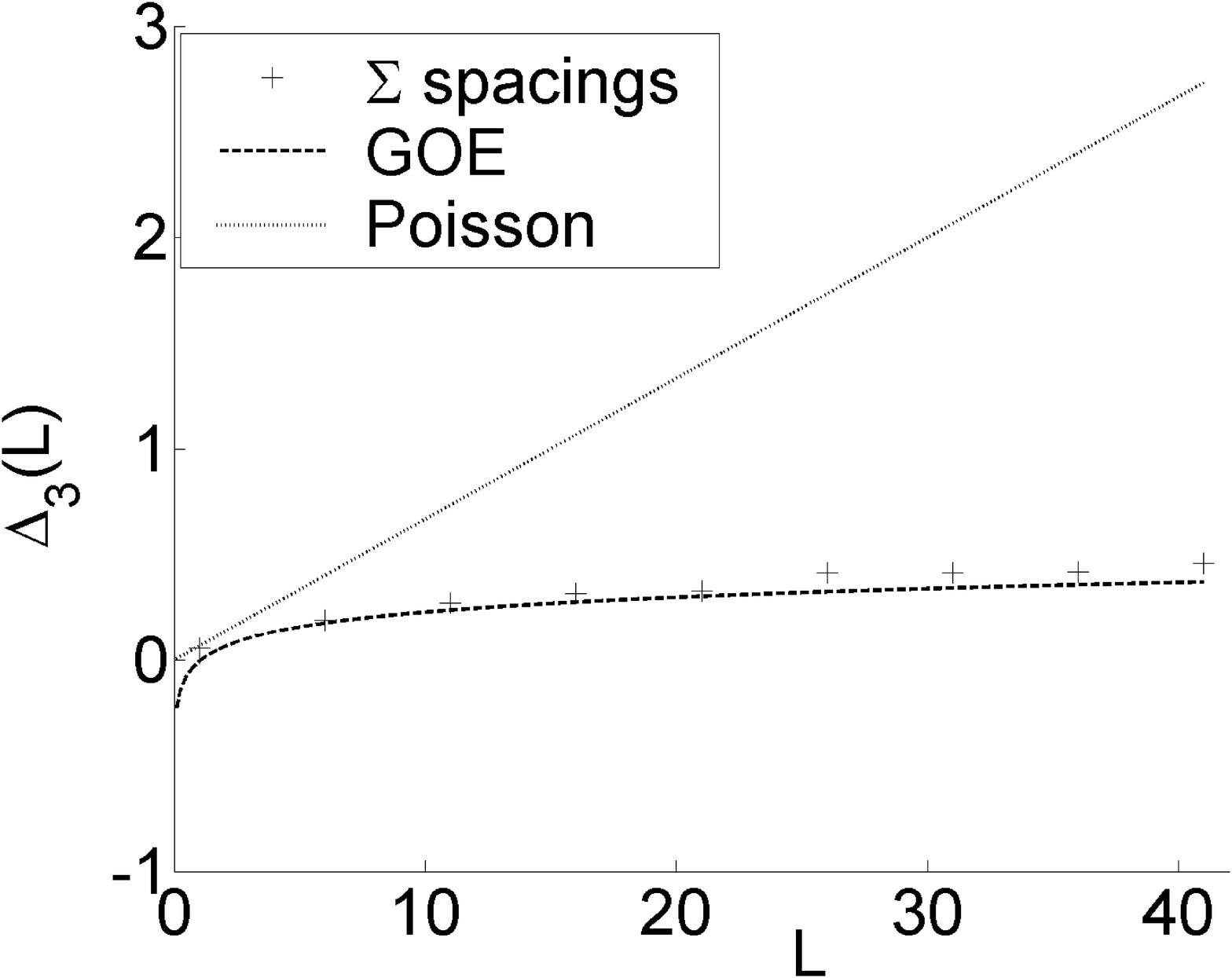}
\end{center}
\vspace{0.0cm}
\caption{
Stadium billiard system in 2-D. Level spacing distribution (top)
from action matrix $\Sigma$ and its spectral 
rigidity $\Delta_{3}$ (bottom).
}
\label{fig:StadiumBilliard}
\end{figure}
\begin{figure}[ht]
\begin{center}
\includegraphics[scale=0.15,angle=0]{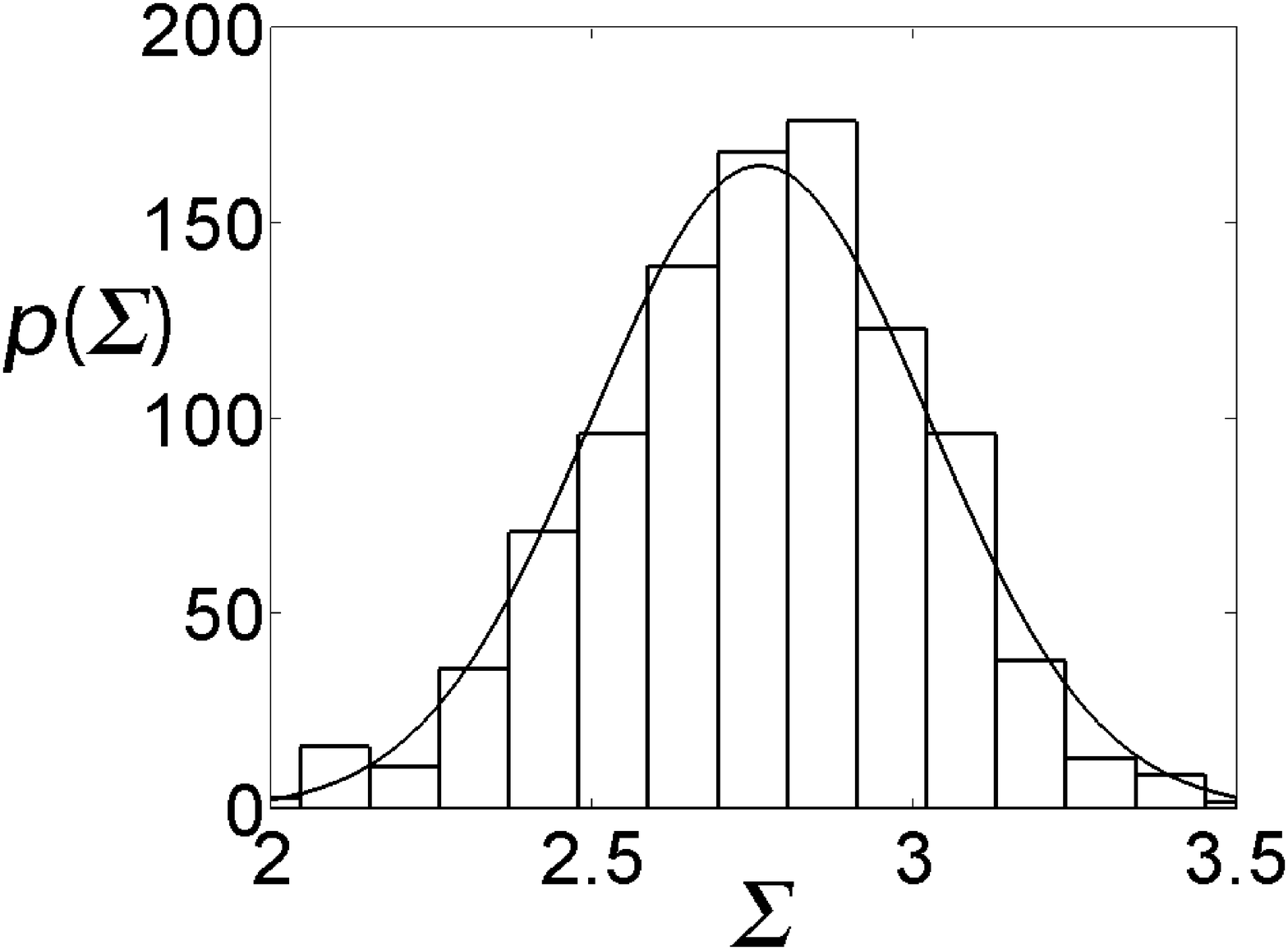}
\includegraphics[scale=0.40,angle=-90]{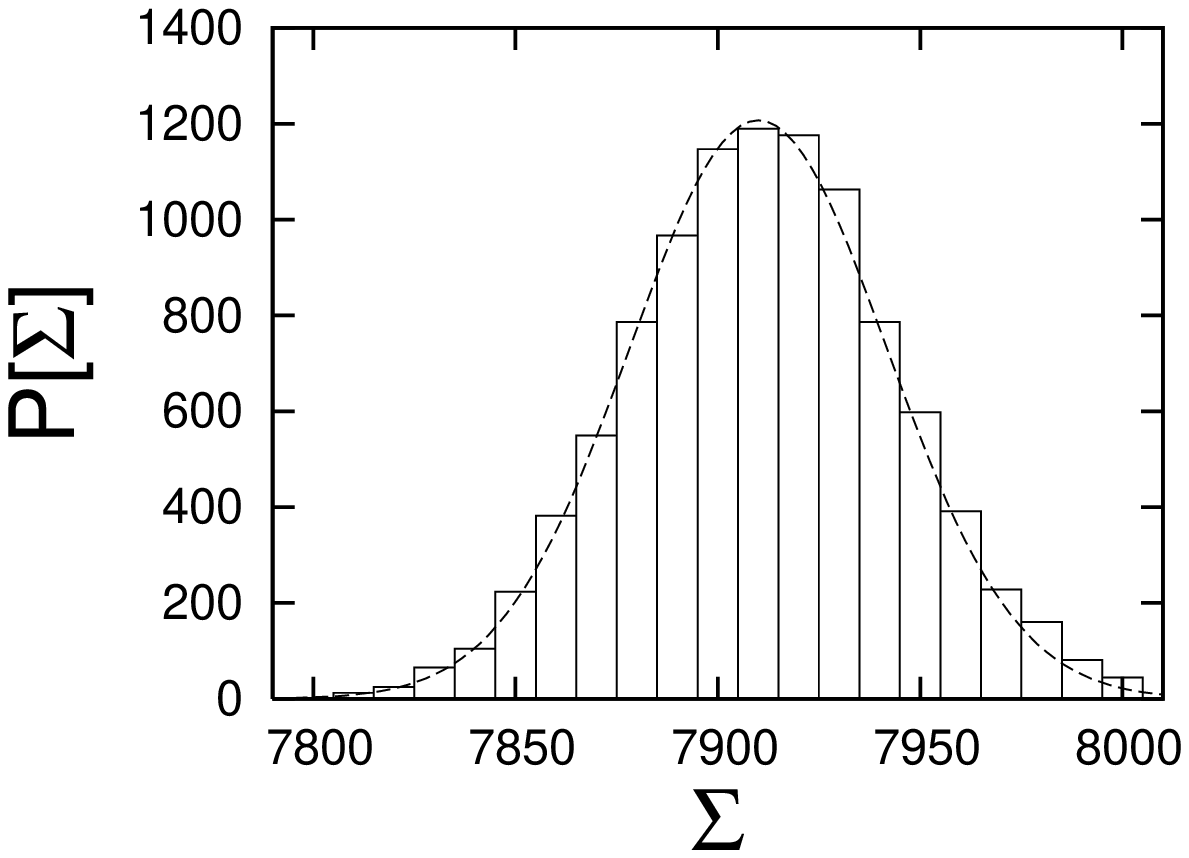}
\end{center}
\caption{
Histrogram of action matrix elements. Top: stadium billiard corresponding to 12 collisions.
Bottom: diffusion (random walk) corresponding to 1000 collisions. A Gaussian behavior is predicted 
by the Central Limit Theorem. 
}
\label{fig:ActionElemDiff+Billiard}
\end{figure}
In quantum systems a discrete level spacing distribution occurs, if the potential supports a bound state spectrum, like in the hydrogen atom. The classical Hamiltonian function $H(x,p,t)$ yields a continuous energy spectrum, and hence is not suitable to play a role like the Hamiltonian of the quantum system. However, the Hamiltonian function $H(x,p,t)$ is closely related to the Lagrange function $L(x,\dot{x},t)$. In order to construct a finite matrix, we consider the classical action $\Sigma$
evaluated along classical trajectories between discrete boundary points, 
\begin{equation}
\label{eq:DefAction}
\Sigma = S[x^{traj}] = 
\int_{o}^{T} dt ~ L(x(t),\dot{x}(t),t)\mid_{x=x^{traj}} ~ .  
\end{equation}
We choose a finite set of discrete points 
$\{x_{1},x_{2},\dots,x_{N}\}$. For any pair of boundary points 
$x_{i},x_{j}$ we compute the corresponding classical trajectory 
$x^{traj}_{x_{i},x_{j}}$ (we assume for the moment that to each 
pair of boundary points corresponds a unique classical trajectory, 
the case of multiple trajectories will be discussed below). 
Then the action matrix is built from matrix elements given by
$\Sigma_{ij} = S[x^{traj}_{x_{i},x_{j}}]$.
We then compute the spectrum of eigenvalues $\sigma_{i}$ of the action 
matrix $\Sigma_{ij}$ and carry out a statistical level spacing 
analysis~\cite{Blumel97,Stockmann99,Haake01}.
Here we want to explore the hypothesis that such action matrix for 
chaotic classical systems plays a role 
analoguous to the Hamiltonian matrix for chaotic quantum systems.

{\it Billiard System}.
Let us consider as example the stadium billiard in 2-D. The billiard 
is symmetric under mirror operation about x- and y-axes at the center. 
In order not to mix different symmetry classes, we consider the billiard 
system (see Fig.[\ref{fig:BilliardScheme}]) with boundary points located 
only in one quarter of the billiard. Trajectories start from and arrive 
at points $x_{i}$ and $x_{j}$, respectively, located at the billiard wall. 
The rule of dynamics is free motion in the 
interior region and perfectly elastic specular collision at the wall. 
Such a classical billiard system is known to be fully chaotic~\cite{Bunimovich79}. 
For a given pair of boundary points, there is an infinite number of trajectories, 
connecting these points. We have discarded trajectories of small angle collisions 
(staying close to the wall). We have classified 
the trajectories according to number of rebounds $N_{reb}$ from the stadium walls. 
One expects chaoticity to increase with $N_{reb}$. However, from the computational 
point of view, $N_{reb}$ can not be too large, because each rebound from the 
curved part of the wall decreases the numerical precision roughly by one order 
of magnitude. The results presented below correspond to 
$N_{reb}$ of the order of 4 to 6, which turned out to be sufficient to 
represent the chaotic behavior. We have taken care that numerical values 
of action matrix elements have a relative error below $10^{-6}$. For the 
computation of trajectories we used a symplectic algorithm. The numerical 
results for the level spacing distribution $P(s)$ of action eigenvalues  
and the spectral rigidity $\Delta_{3}$ are shown in Fig.[\ref{fig:StadiumBilliard}]. 
They represent results of superimposition of spectra corresponding 
to different shapes of the billiard. We considered shapes parametrized 
by horizontal over vertical diameter of the stadium (varying from 2.2 by 
increments of 0.1 to 3.2). For each billiard shape we considered 50 boundary points evenly distributed on the wall of the quarter billiard and computed 
trajectories for all pairs of boundary points. To unfold the spectra we used the technique of Gaussian broadening~\cite{Haake01,Bruus97}. The results are consistent with a Wigner distribution.

{\it GOE behavior of action matrix elements in billiard system}.
In quantum chaos the BGS conjecture postulates equivalence, with 
respect to energy level fluctuations, 
of matrix elements from the (time-reversal invariant) quantum mechanical 
Hamiltonian and GOE random matrix elements. Here we show that classical 
action matrix elements of the stadium billiard behave like 
GOE matrices. We have investigated the statistical behavior of action 
matrix elements for the original stadium billiard. 
The action matrix elements were computed for long trajectories 
(12 rebounds). In GOE all off-diagonal 
matrix elements (and likewise all diagonal matrix elements) obey the same distribution. Thus, for statistical purpose, one can draw values from different matrix elements. Likewise in case of 
the stadium billiard, we computed the action for an 
ensemble of different matrix elements and compiled a histogram. The 
result is shown in Fig.[\ref{fig:ActionElemDiff+Billiard}], and gives 
approximately a Gaussian (however, with non-zero mean). \\
\noindent In order to explain such behavior, we present here an argument 
(not a proof) that this is due to the Central Limit Theorem.
Let us consider the full billiard. The particle (representing the billiard ball) 
alternates between collisions with the semi-circle walls and free motion 
(collisions with straight walls are inessential because the single billiard 
is equivalent to an infinite number of billiard copies attached side-by-side 
at the straight walls and then removing the straight walls). 
Such system is very similar to the system of a particle colliding with 
fixed disks: the so-called 2-D Lorentz gas. Here the particle alternates 
between free motion and collisions with circular disks. Furthermore, the 
Lorentz-gas model of equal disks located on a rectangular grid is equivalent 
to the Sinai billiard. The Lorentz gas model has been shown to have chaotic 
character (positive average Lyapunov exponent) and diffusion character 
(finite diffusion constant)~\cite{Gaspard98}. The Lorentz gas model has 
a different character in the case of dense packing of disks, which, however, 
we do not consider here. In the Lorentz gas model, in the limit of many collisions 
between particle and disks, the particle can be viewed as an isotropic 
random walker. Assuming that a particle moves in fixed time steps $\Delta~ t$, 
at each step a distance $\Delta~ x$ drawn from a uniform probability 
distribution over the interval [0,~5], and travels in a direction of an 
angle drawn from a uniform probability distribution over the interval [0,~2$\pi$], 
the probability density function for the final position of the particle obeys 
the diffusion equation 
$\partial_t\rho=D\nabla^2\rho$ for sufficiently large times. Accordingly, 
the variance in the final position of the particle grows linearly in time, i.e. 
$\langle x^2 \rangle = D~t$ where $D$ is the diffusion constant.
The action of the random walker going from $x_{0}$ to $x_{f}$ is 
\begin{equation}
\Sigma_{0}^{f} = 
\sum_{i=1}^{N} \frac{m}{2} \frac{(\Delta x_{i})^{2}}{\Delta t} ~ ,
\end{equation}
Thus $\Sigma_{0}^{f}$ is a 
sum of random variables, all drawn from the 
same distribution. The central limit theorem implies that $\Sigma_{0}^{f}$ 
is a random variable obeying a Gaussian 
distribution (in the limit $N \to \infty$).  
This is confirmed by a numerical simulation shown in 
Fig.[\ref{fig:ActionElemDiff+Billiard}]. 
Note that the average total action for the isotropic random walker is 
non zero and depends on its mass and mean free path, i.e. the average 
distance travelled in one time step. Since the action 
$\Sigma_{0}^{f}$ of a random walker is distributed according to a 
Gaussian distribution and a particle in a Lorentz gas system presents 
diffusive behavior, we expect that the action 
$\Sigma_{0}^{f}$ of a particle in a Lorentz gas system is distributed 
according to a Gaussian distribution. On the other hand the Lorentz gas 
model is a deterministic classical Hamiltonian system obeying time-reversal symmetry. 
Hence the action matrix elements $\Sigma_{0}^{f}$ form a GOE-type ensemble 
(with non-zero mean value and variance depending on the parameter $N$).
Because GOE random matrices yield a Wignerian 
level spacing distribution~\cite{Mehta91}, we expect the diluted Lorentz gas model, 
in the regime of many collisions, to yield a Wignerian 
level spacing distribution from the classical action matrix.

{\it Harmonic and Anharmonic Oscillator}.
We have also considered the 2-D harmonic oscillator (mass $m$ and 
frequency $\omega$) and also the 2-D anharmonic oscillator (with 
coupling $\lambda x^{2} y^{2}$). 
We have chosen the following parameters: $m=\omega=1$ and $\lambda=8$. 
By computing the Poincar\'e sections (which depend on the energy) we 
have verified that the system is classically almost fully chaotic. 
We superimposed spectra corresponding to variation of 
$\omega$ and $\lambda$, such that $\omega/\lambda=1/8$, and $\lambda$ 
taken from the interval [8,~30]. We computed the action matrix $\Sigma_{ij}$, 
unfolded the spectrum and computed the level spacing 
distribution, as well as the spectral rigidity $\Delta_{3}$.   
The results for the anharmonic oscillator (not shown here) are consistent 
with a Wigner distribution of level spacings  and also consistent with GOE 
behavior in spectral rigidity. In contrast, the harmonic oscillator (integrable 
system) yields a level spacing distribution and spectral rigidity of 
Poissonian nature (results not shown here). In smoothing the raw spectrum and 
unfolding the spectrum, we found it useful to tune the free parameter such 
that to reproduce the auto-correlation coefficient ($C=0$  for Poissonian 
distribution~\cite{Bruus97} and $C=-0.27$ for Wigner distribution~\cite{Bohigas83,Seligman84}). 
Meeting this condition helped to fine-tune the level spacing distribution to 
become closer to Poissonian, or Wignerian, respectively.

{\it Discussion}.
We have suggested to extend random matrix theory, used in chaotic 
quantum systems, to classically chaotic systems. We found for the 
classical harmonic oscillator in 1-D and 2-D (integrable system) 
a Poissonian action level spacing distribution. In contrast, for 
the 2-D classical anharmonic oscillator in the chaotic regime as well 
as for the classical stadium billiard we found a Wignerian action level-spacing 
distribution. \\
\noindent From these findings we are led to propose the following conjecture: 
{\it Spectra from action matrices of time-reversal invariant classical 
K-systems show the same statistical fluctuation 
properties as those predicted by GOE matrices.} 
If this holds, it means that universal laws hold in fully chaotic classical systems.
We have presented a plausibility argument, based on the central limit theorem, 
why action matrix elements of the stadium billiard should obey GOE-type laws. \\
\noindent The diffusion process (random process) and deterministic chaotic motion of the 
Lorentz gas are different with respect to memory: In the random process memory is lost in 
one collision step, while in the chaotic system memory is eradicated exponentially with the number of collisions (given a finite resolution). However, the chaotic Lorentz-gas system, after sufficiently many collisions, is equivalent to a random walk with respect to statistical fluctuation properties. \\
\noindent The stadium billiard, the Sinai billiard and the Lorentz gas 
are ergodic systems~\cite{Bunimovich79}. Ergodicity, i.e. long-time 
averages being equivalent to phase space averages, 
can be seen in the stadium billiard in the following way: the action 
of a  trajectory for sufficiently long time (many collisions with the wall) 
can be decomposed into pieces corresponding to shorter trajectories 
between pairs of boundary points. This represents an average over phase space. 
Each action element follows a Gaussian distribution and due to the 
strong law of large numbers, the total action converges against the mean 
value of the Gaussian. \\  
This work may have the following implications:
(i) Present a new efficient method to measure global chaotic behavior 
in classical systems. 
(ii) Shed light on ergodicity/ergodicity breaking in Hamiltonian flow. 
(iii) Allow a unified description of both, quantum and classical chaos.
(iv) Help understanding why quantum chaos is typically weaker 
than classical chaos (e.g. via an effective quantum action~\cite{Jirari01,Caron01}). 
(v) Help in searching for a proof of Bohigas-Giannoni-Schmit conjecture.

{\it Acknowledgments}.
HK, LJD and KJMM have been supported by NSERC Canada.


\onecolumngrid

\end{document}